\documentclass[showpacs,preprintnumbers,amsmath,amssymb,twocolumn,nofootinbib]{revtex4-1}

\usepackage{subfigure}
\usepackage{graphicx}
\usepackage{dcolumn}
\usepackage{bm}

\begin{document}


\title
{Exact calculations of first-passage properties on the  pseudofractal scale-free web}

\author{Junhao Peng}
\email{pengjh@gzhu.edu.cn}
\affiliation {School of Math and Information Science, Guangzhou University, Guangzhou 510006, China.}

\affiliation {Key Laboratory of Mathematics and Interdisciplinary Sciences of Guangdong
Higher Education Institutes, Guangzhou University, Guangzhou 510006, China}

\author{Elena Agliari}
\email{elena.agliari@mat.uniroma1.it}
\affiliation {Department of Mathematics, Sapienza University, Rome 00185, Italy.}

\author{Zhongzhi Zhang}
\email{zhangzz@fudan.edu.cn}

\affiliation {Shanghai Key Laboratory of Intelligent Information
Processing, School of Computer Science, Fudan University, Shanghai
200433, China}

\date{\today}

\begin{abstract}

In this paper, we consider discrete time random walks  on the pseudofractal scale-free web (PSFW) and we study analytically the related first passage properties. First, we classify the nodes of the PSFW into different levels and propose a method to derive the generation function of the first passage probability from an  arbitrary starting node to the absorbing domain, which is located at one or more nodes of low-level (i.e., nodes with large degree). Then, we  calculate exactly the first passage probability,  the survival probability, the mean and the variance of first passage time by using the generating functions as a tool.  Finally, for some illustrative examples corresponding to given choices of starting node and absorbing domain, we  derive exact and explicit results for such first passage properties. The method we propose can as well address the cases where the absorbing domain is located at one or more nodes of high-level on the PSFW, and it can also be used to calculate the first passage properties on  other networks with self-similar structure, such as  $(u, v)$ flowers and recursive scale-free trees.

\end{abstract}



\maketitle


\textbf{We consider a  recursively-grown complex network, able  to model scale-free networks with small-world effect.  This kind of topology is known to be widespread in biological, social as well as technological systems, and it is often hard to deal with analytically.
\newline
As for the network under investigation (and dynamical processes embedded in it), due to its self-similarity, it is feasible for exact analytical investigations. In particular, we consider the random walk process and we focus on the probability for the walker to first reach a given absorbing domain at a given time. Such a first passage problem underlies many stochastic processes where the
event is triggered by a proper variable reaching a specified value for the first time.
\newline
Most works dealing with random walks and first passage properties just focus on the mean first passage time, which, being just the first moment of
the probability distribution, is not a sufficient measure to fully characterize the first passage dynamics of a system.
\newline
Here, we propose a method to derive the exact generation function of the  first passage probability from an arbitrary starting node to the absorbing domain, and we calculate explicitly the first passage probability, the survival probability, the mean and the variance of the first passage time by using the generating functions as a tool.
\newline
The method we propose can address even more general topologies still exhibiting a self-similar structure.}

\section{Introduction}
\label{intro}

Random walks on complex media have been extensively studied in the past several years~\cite{LO93, Weiss94, HaBe87, Avraham_Havlin04,  MeKl00, ChPe13}. How long does it take a random walker to reach a given absorbing domain (or a trap)? This  time is known in the random walk literature as first passage time (FPT) or trap time~\cite{Redner07, MeyChe11}. Its importance lies in the fact that first passage underlies many stochastic processes in which the event, such as a dinner date, a chemical reaction, the firing of a neuron, or the triggering of a stock option, relies on a variable reaching a specified value for the first time. Therefore FPTs  have generated a considerable amount of work over the last decade~\cite{Condamin05, HeMaKn04, Ki58, Zhsh12, CondaBe07}. A first step consists in the analysis of the mean of this random variable, the mean first-passage time (MFPT).
Noting that the MFPTs are deeply affected by the structural properties of the complex systems~\cite{HaBe87, Avraham_Havlin04}, lots of endeavors have been devoted to derive the exact result of the MFPT to some special nodes and the  MFPT averaged over all the starting nodes (also called mean trap time) on different networks, such as  Sierpinski  gaskets~\cite{KoBa02, BeTuKo10}, Apollonian network~\cite{ZhGuXi09},  Koch networks~\cite{ZhGaxie10},  deterministic recursive trees~\cite{CoMi10,  ZhZhGa10, LiWuZh11, ZhLi11, ZhWu10, ZhLiLin11, Agl08, ZhYu09, WuLiZhCh12,AgCasCatSar15} and some deterministic scale-free networks~\cite{AgBu09, AgBuMa10}. There are also some efforts devoted to exactly calculate the MFPT between any pair of nodes and the  MFPT averaged over all the starting nodes or target nodes on some special trees~\cite{Peng14b,  Peng14c}.

However, the MFPT, being just the first moment of the probability distribution of the FPT, is not a sufficient measure to fully characterize the first passage dynamics of a system. One should further analyze the first passage probability (FPP), i.e., the probability  that the random walker first reaches the absorbing domain at  time $n$ $(0<n<\infty)$. In general, the FPP is deeply affected by the topology of the underlying structure, by the location of the starting site, and by the location of the absorbing domain. Recently, Ref.~\cite{BeChKl10} has presented a method to calculate the asymptotic form of the  FPP between any pair of nodes in confined media. The method can be applied to various models of disordered media, such as fractal networks and percolation clusters. However, when focusing on finite graphs and finite times, the exact calculation of the FPP is still difficult. Up to now,  explicit results for the FPP have been obtained in $1$ dimension,  effectively $1$ dimension geometries and some tree or comb structures, such as Cayley trees, hierarchical trees and
hierarchical   combs.~\cite{Redner07}. But for structures with more complicated typology, to the best of our knowledge, explicit results for the FPP are far more elusive~\cite{BenVo14}.

The pseudofractal scale-free web (PSFW) considered here is  a deterministically growing network which was introduced to model scale-free networks with small-world effect~\cite{Dorogovtsev02}. Due to its self-similarity, the PSFW is feasible for
exact analytical and precise numerical investigations and, in fact, in the past several years, much  effort has been devoted to the study of its properties, such as degree distribution, degree correlation, clustering coefficient~\cite{Dorogovtsev02,zhang07b}, diameter~\cite{zhang07b}, average path length~\cite{zhang07}, and the number of spanning trees~\cite{zhang10}. As for random walks on the PSFW, the MFPTs from any starting node to the hub nodes (i.e., nodes which have maximum degree) and the mean trap time to a given hub node (i.e., the MFPT to the hub node averaged over all the starting nodes) were obtained in Ref.~\cite{ZhQiZh09}. However, the FPP  is still unresolved.

 In this paper, we study discrete time random walks  on the PSFW, aiming at deriving the rigorous solutions of the first passage properties, such as the  first passage probability (FPP), the survival probability and the mean first passage time (MFPT).  First, we classify the nodes of the PSFW into different ``levels'' (briefly, a lower level corresponds to a larger degree, \emph{vide infra}) and propose a method to derive the generating function of the FPP from an  arbitrary starting node to a given absorbing domain. Then, exploiting the generating function tool, we can obtain the FPP, the survival probability, the mean and the variance of the first passage time.

In particular, we focus on some illustrative examples, corresponding to the case that the absorbing domain is located at nodes of low levels. Remarkably, in all these cases, we evidenced that the variance of the first passage time scales quadratically with the mean itself, similarly to what happens in finite-degree self-similar graphs (see e.g., \cite{KahnRed89}).

 This article is organized as follows. Section II presents the network model of the PSFW and some preliminary material on generation function. Section III introduces a convenient labelling of nodes, which shall be useful in the following calculations. Section IV proposes the general method to calculate the first passage properties from an arbitrary node to the absorbing domain. Section V presents the exact and explicit results for the first passage properties in the case the absorbing domain is located at a main hub. Section VI presents the exact and explicit results of the first passage properties in the case there are two absorbing nodes located at two given highly-connected nodes. Finally, Section VII contains conclusions and discussions. Technical and lengthy calculations are collected in the Appendices.

\section{Preliminaries}
\label{sec:1}
\subsection{The pseudofractal scale-free web}
\label{sec:psfw}

\begin{figure}
\begin{center}
\includegraphics[scale=0.6]{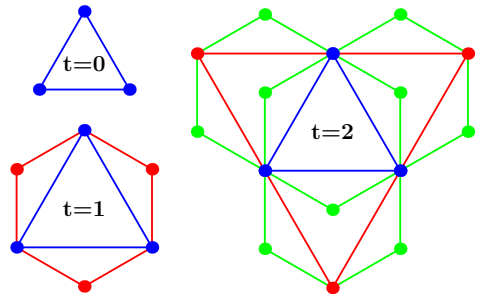}
\caption{The construction of the PSFW with generation $t=0, 1, 2$. At any iteration, each link is replaced by a triangle, in such a way that, for each link, a new node and two new links are introduced. As a consequence, at any iteration, the degree of the existing nodes is doubled.}
\label{fig:1}       
\end{center}
\end{figure}

The PSFW  we studied is  a deterministically growing network  constructed iteratively ~\cite{Dorogovtsev02}. 
 Let $G(t)$ denote the PSFW of generation $t$  ($t\geq 0$). 
  For $t=0$, $G(0)$ is a triangle. For $t\geq 1$, $G(t)$ is obtained from $G(t-1)$: for any edge of $G(t-1)$, a new node is added, which is attached to both the end nodes of
the edge. Fig. \ref{fig:1} shows the construction of the PSFW of generation $t=0$, $1$, $2$.

By construction, it is easy to see that the total number of nodes $V_t$ for $G(t)$ grows exponentially with the number of iterations, being $V_t=(3^{t+1}+3)/2$.

\begin{figure}
\begin{center}
\includegraphics[scale=0.8]{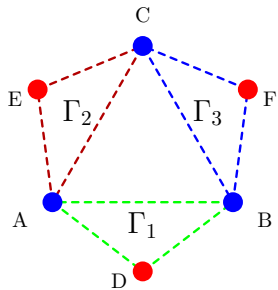}
\caption{Alternative construction of the PSFW which highlights self-similarity: the network of generation $t$, denoted by $G(t)$, is composed of three copies of $G(t-1)$ labeled as $\Gamma_1$, $\Gamma_2$, $\Gamma_3$. Each copy displays three main hubs (say $A, B, D$ in $\Gamma_1$), which are topologically equivalent due to the intrinsic symmetry of the model. Of these main hubs, two are selected (say $A$ and $B$ in $\Gamma_1$) and connected pairwise with the other selected hubs in the other copies. The resulting structure displays $A, B, C$ as main hubs, while nodes labeled by $D$, $E$, $F$, are the second most connected nodes. }
\label{Self_similar}       
\end{center}
\end{figure}

The network also has an equivalent  construction method which highlights its self-similarity~\cite{zhang10, Bobe05}
\footnote{Actually, the PSFW is self-similar in a weak sense:
it contains subgraphs that resemble the whole, but lacks
the affine transformation of scale associated with self-similarity
in fractals. In fact, its dimension is infinite and this is why the name pseudofractal.}.
Referring to Fig. \ref{Self_similar}, in order to obtain $G(t)$, one can make three copies of $G(t-1)$ and join them  at their hubs (i.e., nodes with highest degree)  denoted by $A, B, C$. 

A large class of scale-free networks, built recursively in this way and denoted $(u,v)-flowers$, was introduced in~\cite{RoHa07}; the case analysed here corresponds to the $(1,2)-flower.$

As anticipated, the PSFW is a scale-free network with small-world effect~\cite{Dorogovtsev02, zhang07}. More precisely,
the diameter of the graph increases logarithmically with its size $V_t$ and, for large networks, the number $m(k,t)$ of nodes with degree $k$ decreases as a power of $k$ \cite{Dorogovtsev02}. Actually, the degree spectrum of this graph is discrete, but one can bypass the discreteness of $m(k,t)$ by using the cumulative distribution, hence deriving that the exponent gamma characterizing the degree distribution (hence recovering the expression for the continuous case $P(k) \sim k^{-\gamma}$)  is $\gamma = 1 + \log 3 / \log 2 \approx 2.585$. Remarkably, values of $\gamma \in (2,3)$ are often evidenced in real growing scale-free networks~\cite{Dorogovtsev02}.

We finally notice that the graph exhibits a high clustering coefficient and that it contains numerous loops and hence it is very far from tree-like.

  \subsection{The probability generating function}
\label{sec:PGF}
  Let $\xi$ be a discrete random variable which takes only non-negative integer values, and whose probability distribution is $p_k$ ($k=0,1,2,\cdots$), meaning that the variable $\xi$ assumes value $k$ with probability $p_k$. The probability generating function of $\xi$ is defined as
 \begin{equation}\label{Def_PGF}
    \phi_{\xi}(z)=\sum_{k=0}^{+\infty}z^k p_k.
 \end{equation}
 The probability generating function of $\xi$ is  determined by the probability distribution and, in turn, it uniquely determines the probability distribution. If $\xi_1$ and $\xi_2$ are two random variables with the same probability generating function, then they have the same probability distribution. Given the probability generating function $\phi_{\xi}(z)$ of the random variable $\xi$, we can obtain the probability distribution $p_k$ ($k=0,1,2,\cdots$)  as the coefficient of $z^k$ in the Taylor's series expansion of $\phi_{\xi}(z)$ about $z=0$. This can be restated as $p_k = (\partial^k \phi_{\xi}(z)/ \partial z^k)_{z=0} /k!$.

Also, the $n$-th moment $\langle \xi^n \rangle \equiv \sum_{k=0}^{+\infty} k^n p_k$, can be written in terms of combinations of derivatives (up to the $n$-th order) of $\phi_{\xi}(z)$ calculated in $z=1$. In particular,
\begin{eqnarray}\label{n_moment}
\langle \xi \rangle  &=& \frac{\partial \phi_{\xi}(z)}{ \partial z} \Big |_{z=1},\\
\langle \xi^2 \rangle  &=&  \frac{\partial^2 \phi_{\xi}(z)}{ \partial z^2} \Big|_{z=1} + \frac{\partial \phi_{\xi}(z)}{ \partial z} \Big|_{z=1}.
 \end{eqnarray}

Finally, we list  some properties of the probability generating function~\cite{Rudn04}, which shall be useful in the following:
   \begin{itemize}
  \item Let $\xi_1$ and $\xi_2$ be two independent random variables with probability generating functions $\phi_{\xi_1}(z)$ and $\phi_{\xi_2}(z)$, respectively. Then, the generating function of the distribution of $\xi_1+\xi_2$ reads as
  \begin{equation}\label{Sum_PGF}
    \phi_{\xi_1+\xi_2}(z)=\phi_{\xi_1}(z)\phi_{\xi_2}(z).
 \end{equation}
  \item Let $N$, $\xi_1$, $\xi_2$, $\cdots$ be independent random variables. If $\xi_i$ ($i=1, 2, \cdots$) are identically distributed, each with probability generating function $\phi_{\xi}(z)$, and, being $\phi_{N}(z)$ the probability generating function of $N$, the random variable defined as
 \begin{equation}\label{Sum_PGF}
    S_N=\xi_1+\xi_2+\cdots+\xi_N
 \end{equation}
 has probability generating function
  \begin{equation}\label{Sum_SN}
    \phi_{S_N}(z)=\phi_{N}(\phi_{\xi}(z)).
 \end{equation}
 \end{itemize}

\section{Labelling for the subunits and the nodes of the PSFW}
\label{LEVEL}

As shown in Fig.~\ref{Self_similar}, the  PSFW $G(t)$  is composed of $3$  subunits which are copies  of $G(t-1)$,  and $G(t-1)$ is in turn also composed of $3$ subunits which are copies  of $G(t-2)$. For convenience,  we  classify these subunits  into different levels and let $\Lambda_k$ denote any subunits of level $k$ $(k\geq0)$. More precisely, $G(t)$ is said to be subunit of level $0$, i.e. $\Lambda_0$, then, recursively, for any $k \geq 0$, the three subunits of $\Lambda_k$ are said to be subunits of level $k+1$.
Thus,  $\Lambda_0$ is $G(t)$ itself,  $\Lambda_1$ is a copy of $G(t-1)$, and  $\Lambda_k$ $(k\geq0)$ is a copy of  $G(t-k)$.

Similarly, we classify also the nodes of $G(t)$ into different  levels. For any $k$ $(k\geq0)$,  the  hubs of any subunit $\Lambda_{k}$ are said to belong to level $k$ and nodes belonging to level $k$ also belong to any level $n \geq k$; of course, as $k$ gets larger, the hubs in $\Lambda_k$ display a smaller and smaller degree and, for $k=t$, ``hubs'' only display two neighbours.  Thus, nodes of level $0$ are just the three hubs of $G(t)$, and any node of $G(t)$ belongs to level $t$. 

From another perspective, namely following the building procedure described in Fig.~\ref{fig:1}, the nodes present at iteration $t=0$ belongs to level $0$; at the next iteration, the newly added nodes belong to level $1$, and so on. In this way, nodes present at the $k$-th iteration are said to be of level $k$. As a consequence, nodes belonging to low levels exhibit a high degree.

\begin{figure}
\begin{center}
\includegraphics[scale=0.8]{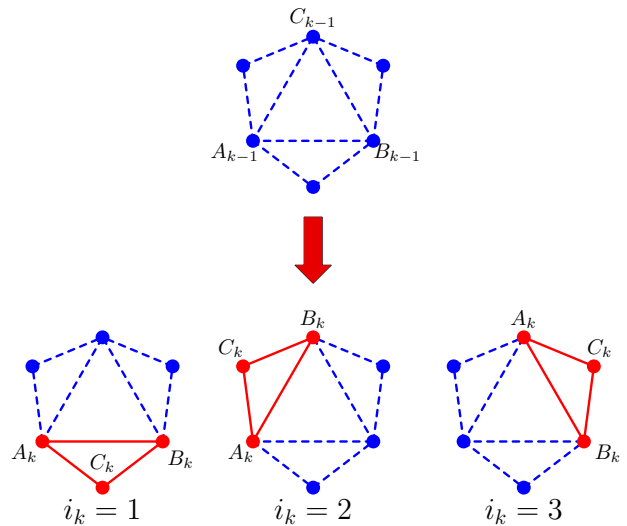}
\caption{The relation  between the value of $i_{k}$ and the location of  subunit $\Lambda_{k}$ in $\Lambda_{k-1}$. Subunit represented by solid red triangle is the  subunit $\Lambda_{k}$ corresponding to value of $i_{k}$ below. The hubs labeled as $A_{k-1}$ and $B_{k-1}$ in $\Lambda_{k-1}$ are also the hubs of $\Lambda_{k}$ labeled as $A_{k}$ and $B_{k}$ while $i_{k}=1$; the hubs labeled as $A_{k-1}$ and $C_{k-1}$ in $\Lambda_{k-1}$ are also the hubs of $\Lambda_{k}$ labeled as $A_{k}$ and $B_{k}$  while $i_{k}=2$; the hubs labeled as $B_{k-1}$ and $C_{k-1}$ in $\Lambda_{k-1}$ are also the  hubs of $\Lambda_{k}$ labeled as $A_{k}$ and $B_{k}$   while $i_{k}=3$.}
\label{sub_k}       
\end{center}
\end{figure}

Now, starting from $G(t)$, for any $k \geq 0$, $3^k$ subunits $\Lambda_k$ are present.
 In order to distinguish similar subunits, analogously to the method of Ref~\cite{MeAgBeVo12}, we label the subunit  $\Lambda_k$ $(1\leq k \leq t)$ by a sequence $\{i_1, i_2, ..., i_{k} \}$, where $i_j \in \{1, 2, 3\}, (1\leq j \leq k)$ labels its location in its parent subunit $\Lambda_{j-1}$. FIG.~\ref{sub_k} shows the construction of $\Lambda_{k-1}$ and the relation  between the value of $i_{k}$ and the location of  subunit $\Lambda_{k}$ in its parent subunit $\Lambda_{k-1}$: all subunit $\Lambda_{k}$ are represented by a  triangle, the one represented by solid red triangle are the  subunit $\Lambda_{k}$ corresponding to value of $i_{k}=1, 2, 3$.

\begin{figure}
\begin{center}
\includegraphics[scale=0.8]{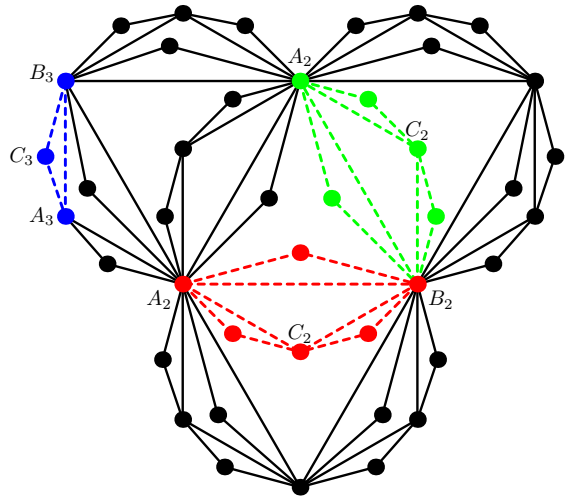}
\caption{The detailed construction of the PSFW with generation $3$. In the PSFW, the structure represented by the dashed blue triangle is a subunit of level $3$ with label $\{2,2,3\}$, its three nodes are labeled by $A_3$, $B_3$, $C_3$;  the structure represented by the dashed red geometry is a subunit of level $2$ with label $\{1,1\}$;  the structure represented by the dashed green geometry is a subunit of level $2$ with label $\{3,1\}$. The hubs of the two subunits of level $2$ are  labeled by $A_2$, $B_2$, $C_2$.}
\label{PSFW_g3}       
\end{center}
\end{figure}

 For convenience, we label the three hubs of  subunit $\Lambda_k$ as $A_k$, $B_k$, $C_k$  and  build the mapping between  hubs of $\Lambda_{k-1}$ and those of $\Lambda_{k}$ as shown in Fig.~\ref{sub_k}.  
 While Fig.~\ref{PSFW_g3} shows the detailed construction of the PSFW with generation $3$ and gives some examples to illustrate the labels of the subunits and their hubs.

\section{The general method}
In this section we derive an exact expression for the generating function of the first passage probability, by focusing on several cases, corresponding to different choices of the starting node and of the absorbing domain.
In particular, the absorbing domain, referred to as $\mathfrak{D}$, is taken to be a subset of the set $\Omega$ including all the nodes of level 1 (i.e. nodes with the largest and the second largest degree).
 In this section the treatment is kept as general as possible, while in the next section we will fix the specific absorbing nodes  and we will derive the explicit expression of the related first passage probability, survival probability, mean and variance of the first passage time.
We stress that these choices are just meant to provide illustrative examples, while the method proposed can be applied in full generality.

\subsection{First passage from a node of level $0$ to any other nodes of level $0$}
\label{FPP0_2}
In this special case the starting node and the absorbing nodes are all main hubs of the PSFW (namely $A, B, C$ in Fig.~\ref{Self_similar}) and it constitutes a reference framework useful for more general cases considered in the following subsections. 

In order to calculate the probability generating function from one hub to any of the other two hubs, we assume that $\mathfrak{D}= \{A, B\}$ and the starting node is $C$. Of course, this choice is completely general due to the symmetry of the underlying structure.

Now, let $\xi(t)$  denote the first passage time from hub $C$ to $\mathfrak{D}$ and let $Q(t,n)$ denote the FPP (i.e., the probability that $\xi(t)=n$).
Then, the generating function for the probability distribution of $\xi(t)$ is:
\begin{equation}\label{Def_psi}
  \theta(t,z)=\sum_{n=0}^{+\infty}z^n Q(t,n),
\end{equation}
where $t$ represents the generation of the PSFW.

As anticipated, we denote with $\Omega$ the set of nodes $\{A,B, C, D, E, F \}$, namely all and only those nodes of level $1$. For any path $\pi$ from $C$ to the absorbing domain, we call $v_j$ the node in $G(t)$ reached at time $j$, in such a way that $\pi=(v_0=C,v_1,\cdots,v_{\xi(t)}=A$ (or $B$)). Also, we introduce the observable $\tau_i=\tau_i(\pi)$, representing the time taken to reach for the $i$-th time any node in $\Omega$ along the path $\pi$. This time can be defined recursively as follows:
\begin{eqnarray}
\tau_0(\pi) &=& 0,\\
\tau_i(\pi) &=&\min\{k: k>\tau_{i-1},v_{\tau_i}\in \Omega, v_{\tau_i}\neq v_{\tau_{i-1}} \},
\end{eqnarray}
Then, we call $N=\min\{i: v_{\tau_i}=A$ (or $B )\}$. Otherwise stated, $N-1$ represents the number of times that the random walker has visited any node in $\Omega \setminus \{A, B\}$, before reaching $A$ or $B$ for the first time. Considering only the set $\Omega$, the path $\pi$ can be restated into a ``simplified path'' defined as
\begin{equation}
\label{Def_simp}
  \sigma(\pi)=(v_{\tau_{0}},v_{\tau_1},\cdots, v_{\tau_N} ).
\end{equation}
The path $\sigma(\pi)$ includes only nodes of level $1$ and the time interval between two steps is stochastic (while the time interval between two steps in the path $\pi$ is trivially one). Thus, $\sigma(\pi)$ is just a path (from $C$ to absorbing hubs) on the PSFW of generation $1$, and $N$ is the first-passage time from hub $C$ to any of the other two hubs on the PSFW of generation $1$. Therefore, the probability generating function of $N$ is $\theta(1,z)$.

Now, let us denote with $\eta_i$ ($i=1,2,\cdots,N$) the time taken to move from $v_{\tau_{i-1}}$ to $v_{\tau_{i}}$, namely $\eta_i=\tau_i-\tau_{i-1}$, which, as highlighted above, is stochastic. In general, $N$, $\eta_1$, $\eta_2$, $\cdots$ are independent random variables and the first-passage time from hub $C$ to the absorbing hubs on $G(t)$ satisfies
\begin{equation}
\label{pathlength}
  \xi(t)= \tau_N - \tau_0 = \eta_1+\eta_2+\cdots+\eta_N.
\end{equation}
Note that nodes $A,B, C, D, E, F $ are hubs of  $\Gamma_1$, $\Gamma_2$, $\Gamma_3$, which are  copies of the PSFW of generation $t-1$. Then $\eta_i$ ($i=1,2,\cdots,N$) are identically distributed random variables, each of them is the first-passage time from hub $C$ to  any of the other two hubs on the PSFW of generation $t-1$. Therefore, the probability generating function of $\eta_i$ ($i=1,2,\cdots,N$) are $\theta(t-1,z)$.

Thus, we can obtain from Eqs.~(\ref{Sum_PGF}), (\ref{Sum_SN}) and (\ref{pathlength}) that the probability generating function of $\xi(t)$ satisfies
\begin{equation}\label{Rec_psi}
  \theta(t,z)=\theta(1,\theta(t-1,z)).
\end{equation}
As for $\theta(1,z)$, we can calculate it through the transition probability matrix for random walks on the PSFW with generation $1$. 
As derived in Appendix~\ref{theta1} (see Eq.~\ref{theta_t1}), $\theta(1,z)= z/(2-z)$, therefore,
\begin{equation}\label{theta_t}
  \theta(t,z)=\frac{z}{2^t - z(2^t-1)}.
\end{equation}

By expanding $\theta(t,z)$ we find
\begin{equation}
Q(t,n) = (2^t-1) \left( 1 - \frac{1}{2^t} \right)^n.
\end{equation}

We can also obtain that the derivative of $\theta(t,z)$ is
\begin{equation}\label{First_d}
  \frac{\partial}{\partial z}\theta(t,z)=\frac{2^t}{[z(2^t-1)-2^t]^2}.
\end{equation}
By posing $z=1$ in Eq.~(\ref{First_d}), we obtain the first moment of FPT, i.e.,
\begin{equation}\label{First_m}
 \langle T_{C \rightarrow \{A,B\}} \rangle =   \langle \xi(t) \rangle =\left.\frac{\partial}{\partial z}\theta(t,z)\right|_{z=1}=2^t,
\end{equation}
where we denoted with $\langle T_{C \rightarrow \{A,B\}} \rangle$ the mean first passage time from $C$ to any node of the absorbing domain $\mathfrak{D} = \{A,B\}$.
Similarly, the second moment reads as
\begin{eqnarray}\label{Second_m}
 \nonumber
 \langle T^2_{C \rightarrow \{A,B\}} \rangle &=&   \left.\frac{\partial^2}{\partial z^2}\theta(t,z)\right|_{z=1} + \left.\frac{\partial}{\partial z}\theta(t,z)\right|_{z=1}\\
 \nonumber
 &=& 2^t(2^{t+1}-1).
\end{eqnarray}
One can see that the FPT $T_{C \rightarrow \{A,B\}}$ scales quadratically with its mean value, suggesting  that the MFPT provides a good estimate only for the expected characteristic order of the first passage time.



\subsection{First passage from a node of level $0$ to the absorbing domain }
\label{GenM_Fpp0}
In this subsection we generalize the results of Sec.~\ref{FPP0_2}  by letting the absorbing domain $\mathfrak{D}$ generic, but still made by a subset of $\Omega=\{A,B, C, D, E, F\}$, namely nodes of level $1$, as shown in  Fig.~\ref{Self_similar}. In the next sections we will treat in detail the cases $\mathfrak{D} = \{A\}$ and $\mathfrak{D} = \{D,E\}$.

Let $\zeta_x(t)$  denote the first passage time from node $x$ to the absorbing  domain and let $P_x(t,n)$ denote the FPP that a random walker, starting at node $x$, first reaches the absorbing domain at  time $n$ (i.e., the probability that $\zeta_x(t)=n$). Then, the corresponding  probability generating  function is defined by:
  \begin{equation}\label{Def_phi}
  \phi_x(t,z)=\sum_{n=0}^{+\infty}z^nP_x(t,n).
\end{equation}
For node $x$ of level $0$, if $x$ is an absorbing node, it is straightforward to see that $\phi_x(t,z)=1$. Thus, we only calculate $\phi_x(t,z)$ for $x$ being a non-absorbing hub. Without loss of generality, we assume that the hub $C$ is non-absorbing node, we obtain
\begin{equation}\label{Rec_phi}
  \phi_C(t,z)=\phi_C(1,\theta(t-1,z)).
\end{equation}
 The detail derivation of Eqs~(\ref{Rec_phi}) is  presented in  Appendix~\ref{App_phi} and $\theta(t-1,z)$ can be calculated from  Eq.~(\ref{theta_t}).

As for $\phi_C(1,z)$, we can calculate it through the transition probability matrix for random walks on the PSFW with generation $1$. 
The detail method are  presented in Appendix~\ref{theta1}  (see Eqs.~\ref{PHIC1}-\ref{PHIC2}),
for different choices of the absorbing set.

 Calculating the first and second order derivative with respect to $z$ on both sides of Eq.~(\ref{Rec_phi}) and letting $z=1$, we obtain
\begin{equation}\label{m1}
  \langle T_{C \rightarrow \mathfrak{D}} \rangle=2^{t-1}\times\left.\frac{\partial}{\partial x}\phi_C(1,x)\right|_{x=1}.
\end{equation}
and
\begin{equation}\label{m2}
  \langle T^2_{C \rightarrow \mathfrak{D}} \rangle = 2^{t}  \left[ \frac{\partial^2}{\partial x^2}\phi_C(1,x) +
2(2^t-1) \frac{\partial}{\partial x}\phi_C(1,x)  \right]_{x=1}, 
\end{equation}
where $\langle T_{x \rightarrow \mathfrak{D}} \rangle$ and $ \langle   T_{x \rightarrow \mathfrak{D} }^2 \rangle$ denote the first and second moments of the FPT from node $x$ to the absorbing domain $\mathfrak{D}$.


\subsection{First passage from an  arbitrary node to the absorbing domain }
\label{Fpp_a1}
In this subsection we generalise the result of Sec.~\ref{GenM_Fpp0}, by letting the starting node be arbitrary. The absorbing domain is still meant to be made up of the nodes in $\mathfrak{D} \subseteq \Omega = \{A, B, C, D, E, F\}$ belonging to level $1$.
We therefore aim to calculate the probability generating  function $\phi_x(t,z)$ which was defined in Eq.~\ref{Def_phi}.

For any subunit $\Lambda_{k}$ ($k\geq0$),  its three hubs are labeled by ${A_k}$, ${B_k}$ and ${C_k}$. As derived in Appendix~\ref{App_g_C_k},
   \begin{eqnarray}\label{g_Ck}
     \phi_{C_k}(t,z)&=&\frac{1}{2}\theta(t-k,z)[\phi_{A_k}(t,z)+\phi_{B_k}(t,z)].
  \end{eqnarray}

Note that we have calculated  $\phi_{A}(t,z)$, $\phi_{B}(t,z)$ and $\phi_{C}(t,z)$ in Sec.~\ref{GenM_Fpp0}, that is to say that $\phi_{A_1}(t,z)$ and $\phi_{B_1}(t,z)$ are obtained for any $i_1=1,2,3$. Posing $k=1$ in Eq.~(\ref{g_Ck}), we gain $\phi_{C_1}(t,z)$ which also shows that $\phi_{A_2}(t,z)$ and $\phi_{B_2}(t,z)$ are obtained for any $i_2=1,2,3$. Using Eq.~(\ref{g_Ck}) recursively, we can calculate the Green function $\phi_{x}(t,z)$ for any node $x$ of the PSFW.

We also find that the first and second moments satisfy the following recursive formulas: 
\begin{eqnarray}\label{T_Ck}
  \langle   T_{C_k \rightarrow \mathfrak{D} } \rangle &=&\frac{1}{2}[  \langle T_{A_k \rightarrow \mathfrak{D}} \rangle +  \langle T_{B_k \rightarrow \mathfrak{D}} \rangle]+2^{t-k}
  \end{eqnarray}
and
  \begin{eqnarray}\label{M2_Ck}
     \langle   T_{C_k \rightarrow \mathfrak{D} }^2 \rangle
      &=&\frac{1}{2}\left[\langle   T_{A_k \rightarrow \mathfrak{D} }^2 \rangle+\langle   T_{B_k \rightarrow \mathfrak{D} }^2 \rangle\right]+2^{2(t-k)+1} \nonumber\\
     &+&2^{t-k}[ \langle T_{A_k \rightarrow \mathfrak{D}} \rangle +  \langle T_{B_k \rightarrow \mathfrak{D}} \rangle]-2^{t-k}.
      \end{eqnarray}
   The detail derivation of Eqs~(\ref{T_Ck}) and (\ref{M2_Ck}) are  presented in Appendix~\ref{App_g_C_k}.

If we can calculate $\langle T_{x \rightarrow \mathfrak{D}}\rangle$ for nodes of level $0$, that is to say that $\langle T_{A_1 \rightarrow \mathfrak{D} }\rangle$ and $\langle T_{B_1 \rightarrow \mathfrak{D}}\rangle$ are obtained for any $i_1=1,2,3$. Let $k=1$ in Eq.~(\ref{T_Ck}), we gain $\langle T_{C_1 \rightarrow \mathfrak{D}}\rangle$ which also shows that $\langle T_{A_2\rightarrow \mathfrak{D}}\rangle$ and $\langle T_{B_2\rightarrow \mathfrak{D}}\rangle$ are obtained for any $i_2=1,2,3$. Using Eq.~(\ref{T_Ck}) recursively, we can calculate  $\langle T_{x \rightarrow \mathfrak{D}}\rangle$ for any node $x$ of the PSFW.

 Similarity, if we can calculate $\langle T_{x \rightarrow \mathfrak{D}}\rangle$ and $\langle T_{x \rightarrow \mathfrak{D}}^2\rangle$ for nodes of level $0$, using Eq.~(\ref{M2_Ck}) recursively, we can calculate  $\langle T_{x \rightarrow \mathfrak{D}}^2\rangle$ for any node $x$ of the PSFW.
\section{First passage properties in the presence of a single absorbing hub}
\label{Exam1}
In this section, we analyse the case of diffusion in a PSFW where one of its main hub, say $A$ (see Fig.~\ref{Self_similar}) is absorbing. We consider in detail two different cases: the random walker starts from another main hub (i.e., either $B$ or $C$) and the random walker starts from an arbitrary node of level $k>0$. For both cases we calculate explicitly the first-passage probability, the survival probability and  the MFPT averaged over starting nodes of level $k$ $(0<k\leq t)$ \footnote{When $k=0$ the starting nodes are two and topologically equivalent, therefore the average over starting nodes simply returns the mean time obtained for any of the two starting nodes.}.

\subsection{First passage properties for a random walker starting from a node of level $0$}
\label{FPP1_0}
Let us assume that the node $A$ is absorbing and let us fix as starting node $C$. We therefore need to calculate $P_C(t,n)$ (of course, by symmetry, $P_B(t,n)=P_C(t,n)$). To this aim, let us recall the result found in Sec.~\ref{GenM_Fpp0} for the generating function $\phi_C(t,z)$ in the presence of a generic boundary set $\mathfrak{D} \subseteq \Omega$, that is $\phi_C(t,z) = \phi_C(1, \theta(t-1,z))$ (see Eq.~\ref{Rec_phi}). Here, $\mathfrak{D}=\{A\}$, for which, as derived in Appendix~\ref{theta1}, $\phi_C(1,z)= z(4 - 3z)$.

Therefore, by merging Eq.~\ref{theta_t} and Eq.~\ref{Rec_phi}, we get
\begin{equation}\label{phic_t}
  \phi_{C}(t,z)= \frac{z}{2^{t+1} - z(2^{t+1}-1)}.
\end{equation}

Now, the first passage probability $ P_C(t,n)$ can be obtained by expanding $ \phi_C(t,z)$ into a  power series of $z$, as
\begin{eqnarray}\label{exp_phict}
  \phi_C(t,z)
             &=&\sum_{n=1}^{+\infty}z^n\frac{1}{2^{t+1}}\left[\frac{2^{t+1}-1}{2^{t+1}}\right]^{n-1}.
\end{eqnarray}
Therefore, the FPP from hub $C$ to the absorbing hub $A$ at time $n$ $(n\geq1)$ is just the coefficient of $z^n$, i.e.,
\begin{eqnarray}\label{FPP_C}
 P_C(t,n)&=&\frac{1}{2^{t+1}}\left(1 - \frac{1}{2^{t+1}}\right)^{n-1}, 
\end{eqnarray}
which shows that the FPP decreases exponentially with the number of steps.
In fact, in the limit of large size (i.e., $t \gg 1$), we can reshuffle this expression as $ P_C(t,n) \approx 2^{-(t+1)}\exp[-(n-1)/2^{t+1}]$, suggesting that the characteristic time for the event to occur scales as $2^{t+1}$.


We can also obtain the survival probability (i.e., the probability that a random walker has not hit the absorbing hub $A$ by time $n$) as
\begin{eqnarray}\label{FPP_C}
 S_C(t,n)&=&1-\sum_{k=0}^{n}P_C(t,k) \nonumber \\
             &=&\left(1- \frac{1}{2^{t+1}}\right)^{n} 
             .
\end{eqnarray}
Therefore, the survival probability converges exponentially to zero, as expected for finite size structures.

As for the MFPT, it is easy to derive that
\begin{equation}
 \frac{\partial}{\partial z}\phi_C(1,z)=\frac{4}{[3z-4]^2}.
\end{equation}
Thus, recalling Eq.~(\ref{m1}), the MFPT from $C$ to $A$ reads as
\begin{equation}\label{TC_C1}
\langle  T_{C \rightarrow \{A\}} \rangle= 2^{t-1} \times \left.\frac{\partial}{\partial x}\phi_C(1,x)\right|_{x=1}=2^{t+1}.
\end{equation}

Since the volume of the underlying structure scales as $V_t \sim 3^t$ for large sizes, the previous expression can be restated as
\begin{equation}\label{TC_C1b}
\langle  T_{C \rightarrow \{A\}} \rangle= V_t^{\log_3 2} = V_t^{1/(\gamma-1)},
\end{equation}
namely the MFPT scales sublinearly with the volume ($ 1/(\gamma-1) \approx 0.63$). Of course, such a fast dynamics is due to the centrality of the absorbing site.

The second moment turns out to be
 \begin{eqnarray}\label{TC_C1_2nd}
\langle  T^2_{C \rightarrow \{A\}} \rangle &=& \left.\frac{\partial^2}{\partial z^2}\phi_C(t,x)\right|_{z=1} + \left.\frac{\partial}{\partial z}\phi_C(t,x)\right|_{z=1}\\
&=&2^{t+1}(2^{t+2}-1).
\end{eqnarray}
Therefore, again, we have that the standard deviation of the first passage time scales linearly with the mean.

This calculation can be extended to higher moments, leading to $\langle  T^q_{C \rightarrow \{A\}} \rangle \sim \langle  T_{C \rightarrow \{A\}} \rangle^q$, for finite $q \in \mathbb{N}$, consistently with the result found in \cite{KahnRed89} for exact self-similar structures with finite degree.

\subsection{First passage properties for a random walker starting from an arbitrary node of level $k$  }
\label{FPP1_a}
Besides nodes of level $0$, we can also derive the first-passage properties for a random walker starting from some nodes of level $k$ ($0<k\leq t$), still keeping as absorbing domain $\mathfrak{D} = \{A\}$.

For example, we derive the first-passage properties for a random walker starting from the hub $C_k$ ($0<k\leq t$) of subunit $\Lambda_k$  labeled by $\{i_1, i_2, ..., i_{k}\}=\{\underbrace{1, 1,\cdots, 1}_k\}$ (e.g.,  Fig.~\ref{PSFW_g3} shows the subunit labeled by $\{1,1\}$ in the PSFW of generation $3$).

Note that $A_{j}\equiv A_{j-1}$, $B_{j}\equiv B_{j-1} $ while $i_{j}=1$. Thus, if $\{i_1, i_2, ..., i_{k}\}=\{\underbrace{1, 1,\cdots, 1}_k\}$, we have
\begin{eqnarray} \label{mappingk1}
             \begin{array}{ll}
   A_{j}\equiv A_{0}\equiv A, B_{j}\equiv B_{0} \equiv B,  & j=1,2,\cdots, k.
   \end{array}
\end{eqnarray}
Substituting $\theta(t-k,z)$  from Eq.~(\ref{theta_t}) in Eq.~(\ref{g_Ck}),
   \begin{eqnarray}\label{ph_Ck1}
     \phi_{C_k}(t,z)
     &=&\frac{\frac{1}{2}z}{2^{t\!-\!k} - z(2^{t\!-\!k}\!-\!1)}\nonumber \\
     &&+\frac{\frac{1}{2}z^2}{[2^{t\!-\!k} - z(2^{t\!-\!k}\!-\!1)][2^{t\!+\!1} - z(2^{t\!+\!1}\!-\!1)]}.
  \end{eqnarray}

It is worth noting that
\begin{eqnarray}\label{part1}
  \frac{\frac{1}{2}z}{2^{t-k} - z(2^{t-k}-1)}=\sum_{n=1}^{+\infty}z^n\frac{1}{2^{t-k+1}}\left[\frac{2^{t-k}-1}{2^{t-k}}\right]^{n-1},\nonumber \\
\end{eqnarray}
and that
   \begin{eqnarray}\label{part2}
     &&\frac{\frac{1}{2}z^2}{[2^{t\!-\!k} - z(2^{t\!-\!k}\!-\!1)][2^{t\!+1} - z(2^{t\!+\!1}\!-\!1)]} \nonumber \\
     &=&\frac{z^2}{2^{2t-k+2}}\left\{\frac{2^{k+1}-2^{t+1}}{2^{k+1}-1}\sum_{n=0}^{+\infty}z^n\left[\frac{2^{t-k}-1}{2^{t-k}}\right]^n \right.\nonumber \\
     &&\left.+\frac{2^{t\!+\!1}\!-\!1}{2^{k+1}\!-\!1}\sum_{n=0}^{+\infty}z^n\left[\frac{2^{t+1}\!-\!1}{2^{t+1}}\right]^n \right\}.
  \end{eqnarray}
  Inserting Eqs.~(\ref{part1}) and (\ref{part2}) into in Eq.~(\ref{ph_Ck1}) and calculating the coefficient of $z^n$, we obtain the FPP from $C_k$ to the absorbing hub $A$, i.e. 
  \begin{eqnarray}\label{FPP_Ck1}
P_{C_k}(t,n)
&=&\frac{1}{ 2^{t-k} (2^{k+1}-1)} \\
\nonumber
&\times& \left [ (2^{k}-1) \left(1- \frac{1}{2^{t-k}} \right)^{n-1} +\frac{1}{2} \left( 1- \frac{1}{2^{t+1}} \right)^{n-1}  \right ]
\end{eqnarray}

Therefore, the survival probability is
\begin{eqnarray}\label{FPP_C}
 S_{C_k}(&t&,n)=1-\sum_{i=0}^{n}P_{C_k}(t,i) \nonumber \\
            &\!=\!&\frac{{ \!-\!1}}{{2^{k+1}\!-\!1}}\left[1 - \frac{1}{2^{t\!-\!k}}\right]^{n}\!+\!\frac{{2^{k+1}}}{{2^{k+1}\!-\!1}}\left[1 - \frac{1}{2^{t\!+\!1}}\right]^{n}, 
\end{eqnarray}
and it is vanishing exponentially with time, as expected.

Calculating the derivative of $\phi_{C_k}(t,z)$ with respect to $z$ and posing $z=1$, we obtain the MFPT from $C_k$ to the absorbing hub $A$ as
\begin{equation}\label{TCk_C1}
 \langle T_{C_k \rightarrow \{A\}} \rangle=\left.\frac{\partial}{\partial z}\phi_{C_k}(t,z)\right|_{z=1}=2^{t}(1 + 2^{-k}).
\end{equation}
%
Notice that for low degree nodes, corresponding to large $k$, the MFPT is smaller. In fact, starting from these nodes, it is less likely to ``get lost'' without ever reaching the hub. Also,  $\langle T_{C_k \rightarrow \{A\}} \rangle$ still scales sub linearly with the volume size.

The second moment reads as
\begin{eqnarray}\label{TCk_C1_2nd}
\nonumber
 \langle T^2_{C_k \rightarrow \{A\}} \rangle &=& \left.\frac{\partial^2}{\partial z^2}\phi_{C_k}(t,z)\right|_{z=1} + \left.\frac{\partial}{\partial z}\phi_{C_k}(t,z)\right|_{z=1} \\
 &=& 2^t [2^{t+1}(2 + 2^{-k} + 2^{-2k}) -1 - 2^{-k}].
\end{eqnarray}
The variance turns out to be $\langle T^2_{C_k \rightarrow \{A\}} \rangle -   \langle T_{C_k \rightarrow \{A\}} \rangle^2 = 2^{2t}(3 + 2^{-2k}) -  \langle T_{C_k \rightarrow \{A\}} \rangle$, in such a way that, in the limit of large size, the ratio $\sqrt{\langle T^2_{C_k \rightarrow \{A\}} \rangle -   \langle T_{C_k \rightarrow \{A\}} \rangle^2}/  \langle T_{C_k \rightarrow \{A\}} \rangle$ grows with $k$ and, for large $k$, it saturates to $\sqrt{3}$. Therefore, the estimate provided by the MFPT is worse when $C_k$ is a low-degree node pertaining to peripheral subunit.

 \subsection{ MFPT averaged over  starting nodes of level $k$}
 \label{MTT1}
Since nodes of level $k$ are, by definition, the hubs of subunits of level $k$, they can be labeled as $A_k$, $B_k$, and $C_k$ in each subunit. For convenience, let
\begin{equation}\label{Def_TK}
  \mathcal{T}^{(k)}\equiv \left( \begin{array}{c} \langle T_{A_k \rightarrow \mathfrak{D}} \rangle \\\langle T_{B_k \rightarrow \mathfrak{D}} \rangle \\\langle T_{C_k \rightarrow \mathfrak{D}} \rangle \end{array} \right).
\end{equation}
In order to calculate the MFPT averaged over all staring nodes of level $k$ ($k>0)$, we must calculate the sum of the MFPT over all starting nodes of level $k$. Since each subunit of level $k$ is in one to one correspondence with a sequence  $\{i_1,\cdots,i_k\}$,  one has to calculate $\sum_{\{i_1,\cdots,i_k\}} \mathcal{T}^{(k)}$. Let
  \begin{equation}\label{sigma_k}
  \Sigma_k=\sum_{j=1}^3 \left( \sum_{\{i_1,\cdots,i_k\}}\mathcal{T}^{(k)} \right)_j,
\end{equation}
where the second summation is run over all the subunits of  level $k$ (i.e., let $\{i_1,\cdots,i_k\}$ run over all the possible values), and the first summation just adds the three entries of the vector $\sum_{\{i_1,\cdots,i_k\}} \mathcal{T}^{(k)}$ together. As derived in Appendix~\ref{sum_c1},
 \begin{eqnarray}\label{SSTC}
     \Sigma_k =3^{k}(2^{t+2}+2^t-2^{t-k}).
  \end{eqnarray}
But the MFPT $\langle T_{x \rightarrow \{A\}} \rangle$ for each starting node $x$ of level $i$ $(0<i\leq k)$  appears many times in  $\Sigma_k$.
 We must take into account the repetitions of $\langle T_{x \rightarrow \{A\}} \rangle$  in $\Sigma_k$ when we calculate the average of the MFPT over all the starting nodes of level $k$. We find that $\langle T_{x \rightarrow \{A\}} \rangle$  for each starting node $x$ of level $i$ $(0<i\leq k)$  appears $2^{k-i}$ times in  $\Sigma_k$ and it appears just once in $\Sigma_k-\sum_{i=0}^{k-1}\Sigma_i$. Therefore, the average of the MFPT over all the starting nodes of level $k$ is
\begin{eqnarray}\label{MTTk}
\overline{ \langle T \rangle}^{(k)}
   &=&\frac{\Sigma_k-\sum_{i=0}^{k-1}\Sigma_i}{V_k-1}\nonumber \\
   &=& 2^{t} \left[ 1 + \frac{2}{3} \times \frac{1 + 2^{-k}}{1 + 3^{-k-1}} \right] .
\end{eqnarray}
By posing $k=t$ in Eq.~(\ref{MTTk}), we obtain the MFPT averaged over all the starting nodes of PSFW. The result  is consistent with that obtained in Ref.~\cite{ZhQiZh09} and  the correctness of our method is verified.

Since the volume of the underlying structure scales as $V_t \sim 3^t$ for large sizes, Eq.~(\ref{MTTk}) can be restated as
\begin{equation}\label{MTTka}
\overline{ \langle T \rangle}^{(k)} \approx V_t^{1/(\gamma-1)}\left[ \frac{5}{6}+ \frac{1}{3\times 2^{k}}+\frac{1}{2\times3^{k+1}}\right],
\end{equation}
which shows that $\overline{ \langle T \rangle}^{(k)}$  also scales sublinearly with the volume, but it decreases with $k$.

\section{First passage properties in the presence of two absorbing nodes}
\label{Exam2}
In this section, we assume that $\mathfrak{D} = \{D,E\}$, namely that only the two nodes labeled by $D$ and $E$, as shown in Fig. \ref{Self_similar},  are absorbing nodes. Similar to Sec.~\ref{Exam1}, we calculate the first-passage probability, the survival probability and  the MFPT averaged over starting nodes  of level $k$ $(0<k\leq t)$.

\subsection{First passage properties for a random walker starting from a node of level $0$ }
\label{FPP2_0}
In this case, all the three nodes of level $0$ (i.e. the three hubs $A$, $B$, $C$ of the PSFW) are non-absorbing hub. By symmetry, $\phi_C(t,z)=\phi_B(t,z)$. Thus, we only need to calculate $\phi_A(t,z)$ and $\phi_C(t,z)$. As derived in Appendix~\ref{theta1}, $\phi_A(1,z)=[z(2-z)]/(4-3z)$ (see Eq.~\ref{PHIA2}) and $\phi_C(1,z)= z/(4-3z)$ (see Eq.~\ref{PHIC2}). Therefore
\begin{eqnarray}\label{phia_t2}%
  \phi_A(t,z)&=&\phi_A(1,\theta(t-1,z)) \nonumber \\
  &=&\frac{z[2^t\!-\!z(2^t-1)]}{[2^{t\!-\!1} - z(2^{t\!-\!1}\!-\!1)][2^{t\!+\!1} - z(2^{t\!+\!1}\!-\!1)\!]},%
\end{eqnarray}
and
\begin{equation}\label{phic_t2}
  \phi_C(t,z)=\phi_C(1,\theta_{t-1}(z))=\frac{z}{2^{t+1} - z(2^{t+1}-1)}.
\end{equation}
We find that $\phi_C(t,z)$ is the same as the result of Sec.~\ref{FPP1_0}, namely the FPP (and all the related properties) from hub $C$ to $\mathfrak{D} = \{ D, E\}$ and from hub $C$ to $\mathfrak{D} = \{ A \}$ are the same.

Expanding $ \phi_A(t,z)$  into  power series of $z$,
\begin{eqnarray}\label{exp_phict}
  \phi_A(t,z)
             &=&\sum_{n=1}^{+\infty}\frac{ z^n}{3\times2^{t}}\left\{2\left[\frac{2^{t\!-\!1}\!-\!1}{2^{t\!-\!1}}\right]^{n\!-\!1}\!+\!\left[\frac{2^{t\!+\!1}\!-\!1}{2^{t\!+\!1}}\right]^{n\!-\!1}\right\},  \nonumber
\end{eqnarray}
and the FPP from hub $A$ to the absorbing domain at time $n$ $(n\geq1)$ is just the coefficient of $z^n$, i.e.,
\begin{equation}\label{FPP_C}
 P_A(t,n)=\frac{ 1}{3\times2^{t}}\left [ 2\left(\frac{2^{t\!-\!1}\!-\!1}{2^{t\!-\!1}}\right)^{n\!-\!1}\!+\!\left(\frac{2^{t\!+\!1}\!-\!1}{2^{t\!+\!1}}\right)^{n\!-\!1}\right]  .
\end{equation}
The survival probability  is
\begin{eqnarray}\label{FPP_C}
 S_A(t,n)&=&1-\sum_{k=0}^{n}P_A(t,k) \nonumber \\
               &=&\frac{1}{3}\left(1 - \frac{1}{2^{t-1}}\right)^{n}+\frac{2}{3}\left(1 - \frac{1}{2^{t+1}}\right)^{n}.
\end{eqnarray}
Moreover, it is easy to derive that
\begin{equation}
 \frac{\partial}{\partial z}\phi_A(1,z)=(2z-2)(3z-4)^{-1}+z(z-2)(3z-4)^{-2}.
\end{equation}
Thus, recalling Eq.~\ref{m1}, the MFPT from hub $A$ to the absorbing domain is
\begin{equation}\label{TA_C2}
 \langle T_{A \rightarrow \{D,E\}} \rangle=2^{t-1}\times\left.\frac{\partial}{\partial x}\phi_A(1,x)\right|_{x=1}=3\times2^{t-1},
\end{equation}
and, again, it scales sublinearly with the volume.

The variance reads as
\begin{eqnarray}\label{TA_C2_2nd}
\nonumber
 \langle T^2_{A \rightarrow \{D,E\}} \rangle &=& \left.\frac{\partial^2}{\partial z^2} \phi_A(z,x)\right|_{z=1} + \left.\frac{\partial}{\partial z} \phi_A(z,x)\right|_{z=1}\\
 &=& 2^{t-1}(11 \times 2^t +3).
\end{eqnarray}
Also in this case, the estimate provided by the mean $\langle T_{A \rightarrow \{D,E\}}\rangle $ scales linearly with the related standard deviation.

\subsection{First passage properties for a random walker starting from  an arbitrary node of level $k$ }
\label{FPP2_a}
In general, being $\mathfrak{D} = \{D,E\}$ we can also derive the first-passage properties for a random walker starting from any arbitrary node of level $k$ ($0<k\leq t$). For example, we derive the first-passage properties for a random walker starting from the hub $C_k$ ($0<k\leq t$) of subunit $\Lambda_k$  labeled by $\{i_1, i_2, ..., i_{k}\}=\{\underbrace{3, 1,\cdots, 1}_k\}$ (e.g.,  Fig.~\ref{PSFW_g3} shows the subunit labeled by $\{3,1\}$ in the PSFW of generation $3$).
Note the following mappings for this case
\begin{eqnarray} \label{mappingk2}
  \begin{array}{ll}
   A_{j}\equiv  C, B_{j}\equiv  B,  & j=1,2,\cdots, k.
   \end{array}
\end{eqnarray}

Similar to the derivation of Eq.~(\ref{ph_Ck1}), we have 
   \begin{eqnarray}\label{ph_Ck2}
     \phi_{C_k}(t,z)
     &=&\frac{-\frac{1}{2}z}{z(2^{t-k}-1)-2^{t-k}}[\phi_{B}(t,z)+\phi_{C}(t,z)]\nonumber \\
     &=&\frac{z^2}{[2^{t -k} - z(2^{t\!-\!k}-\!1)] [2^{t\!+\!1} - z(2^{t\!+\!1}\!-\!1)]}.
  \end{eqnarray}
Similar to  the derivation of Eq.~(\ref{FPP_Ck1}),
   expanding $\phi_{C_k}(t,z)$ into a power series of $z$ and   calculating the coefficient of $z^n$, we obtain the FPP from $C_k$ to the absorbing domain, i.e. 
  \begin{eqnarray}\label{FPP_Ck2}
 P_{C_k}(t,n)
%
&=&\frac{1}{2^{t\!+\!1}\!-\!2^{t\!-\!k}} \left[   \left(1 -\frac{1}{2^{t\!-\!k}}\right)^{n\!-\!1} + \left(1 - \frac{1}{2^{t\!+\!1}}\right)^{n\!-\!1} \right].
\end{eqnarray}

Therefore, the survival probability is
\begin{eqnarray}\label{FPP_C}
 S_{C_k}(t,n)&=&1-\sum_{i=0}^{n}P_{C_k}(t,i) \nonumber \\
            &=&\frac{{2^{k}\!-\!1}}{{2^{k+1}\!-\!1}}\left[1 - \frac{1}{2^{t\!-\!k}}\right]^{n}\nonumber \\
            &\!+\!&\frac{{2^{k}}}{{2^{k+1}\!-\!1}}\left[1- \frac{1}{2^{t\!+\!1}}\right]^{n}. 
\end{eqnarray}

Following the mappings shown as Eq.~(\ref{mappingk2}), we get $\langle T_{A_k \rightarrow \{D,E\} } \rangle=\langle T_{C \rightarrow \{D,E\}} \rangle$ and $\langle T_{B_k \rightarrow \{D,E\}} \rangle=\langle T_{B \rightarrow \{D,E\}} \rangle$, hence, replacing $\langle T_{B \rightarrow \{D,E\}}\rangle$ and $\langle T_{C \rightarrow \{D,E\}} \rangle$ from Eq.~(\ref{TC_C1}) in Eq.(\ref{T_Ck}), we obtain
 \begin{eqnarray}\label{T_Ck_c2}
     \langle T_{C_k \rightarrow \{D,E\}} \rangle &=&\frac{1}{2}[T_{A_k \rightarrow \{D,E\} }+T_{B_k \rightarrow \{D,E\}}]+2^{t-k}\nonumber \\
     &=&2^{t+1}+2^{t-k}.
  \end{eqnarray}

 \subsection{MFPT averaged over  starting nodes of level $k$}
  \label{MTT2}

  In the case that nodes $D$, $E$ are absorbing nodes, the method for calculating the MFPT averaged over  starting nodes of level $k$ is similar to the case where $A$ is the absorbing node. As derived in Sec.~\ref{FPP2_0}, $\langle T_{A \rightarrow \{D,E\}} \rangle=3\cdot2^{t-1}$, $\langle T_{B \rightarrow \{D,E\}} \rangle =\langle T_{C\rightarrow \{D,E\}} \rangle=2^{t+1}$.  Therefore
  $$\Sigma_0=\langle T_{A \rightarrow \{D,E\}} \rangle+\langle T_{B \rightarrow \{D,E\}} \rangle+\langle T_{C \rightarrow \{D,E\}} \rangle=11\times2^{t-1}.$$
   For $0<k\leq t$, as derived in Appendix~\ref{sum_c2},
   \begin{eqnarray}\label{SSTC2}
\Sigma_k
      =3^{k}(2^{t+2}+2^t-2^{t-k}).
  \end{eqnarray}
Therefore, the average of the MFPT over all the starting nodes of level $k$ $(0<k\leq t)$ is
\begin{eqnarray}\label{MTTk2}
\overline{ \langle T \rangle}^{(k)}
   &=&\frac{\Sigma_k-\sum_{i=0}^{k-1}\Sigma_i}{V_k-2}\nonumber \\
   &=& 2^{t+1}\frac{\frac{5}{2}\cdot3^{k}+ \left(\frac{3}{2}\right)^k-1}{3^{k+1}-1}\nonumber \\
   &\approx& V_t^{1/(\gamma-1)}\left[ \frac{5}{6}+ \frac{1}{3\times2^{k}}-\frac{1}{3^{k+1}}\right].
\end{eqnarray}
Also in this case,  $\overline{ \langle T \rangle}^{(k)}$  scales sublinearly with the volume and decreases with $k$.

\section{Conclusions}
\label{sec:4}

We have proposed a general method to calculate exactly the first passage probability for random walks on the PSFW in the presence of an absorbing domain located at some nodes of high coordination. From the knowledge of the first passage probability one can derive a detailed description of the problem, getting, for instance, the survival probability, the mean and the variance of first passage time. We calculated explicitly results of the first passage properties for some illustrative examples, corresponding to the case that the absorbing domain is located at one of the three main hubs and to the case that there are two absorbing nodes located at two nodes among those with the second largest degree.
In all these cases we evidenced that the variance of the first passage time scales quadratically with the mean itself.

Of course, the method proposed here is also suitable for other cases where the absorbing domain is located at one or more nodes of low-degree on the PSFW. The method can also be used on other self-similar graph such as $(u, v)$ flower, T-graph, recursive fractal scale-free trees, the recursive non-fractal scale-free trees and etc.

\begin{acknowledgments}
This work was supported  by the scientific research program of Guangzhou municipal colleges and universities under Grant No. 2012A022. ZZZ was supported by the National Natural Science Foundation of China under Grant No. 11275049.
\end{acknowledgments}

\appendix
\section{PGF of FPP on the PSFW of generation $1$  }
\label{theta1}
The PSFW of generation $1$, has just $6$ nodes.
Let
 $$\Pi=(P_{xy})_{6\times6}$$
  be the  transition probability matrix for random walks on the PSFW of generation $1$. This means
\begin{equation}
\label{Pxy}
  P_{xy}=\left\{ \begin{array}{ll} \frac{1}{d_x} & \text{if $x\sim y$, and $x$ is not an absorbing node}\\ 0 & \textrm{others} \end{array} \right.,
\end{equation}
where $x\sim y$ means that there is an edge between $x$ and $y$ and $d_x$ is the degree of node $x$.

Therefore we can calculate the probability generating function of the first passage time directly by
 \begin{eqnarray}\label{Formular_PGF}
\Phi(z)=\sum_{n=0}^{+\infty}(z\Pi)^n=(I-z\Pi)^{-1},
\end{eqnarray}
where $\Phi(z)=(\phi_{xy}(z))_{6\times6}$ and  $\phi_{xy}(z)$ is the probability generating function of passage time from node $x$ to $y$. But if $y$ is an  absorbing node, $\phi_{xy}(z)$ is just the probability generating function of first passage time from node $x$ to $y$.

\paragraph*{Exact calculation of  $\theta(1,z)$.} Let two hubs $B$ and $C$ be the absorbing nodes. Then, all coordinates of the second and the third row of the  transition probability matrix $\Pi$ are assigned $0$. That is
\begin{equation}
\label{PM1}
  \Pi=\left( \begin{array}{llllll}
  0 & \frac{1}{4} & \frac{1}{4}  & \frac{1}{4} & \frac{1}{4} &0 \\
  0 & 0 & 0  & 0 & 0 & 0 \\
  0 & 0 & 0  & 0 & 0 & 0 \\
   \frac{1}{2} & \frac{1}{2} & 0  & 0 & 0 & 0 \\
    \frac{1}{2} & 0 & \frac{1}{2}  & 0 & 0 & 0 \\
     0 & \frac{1}{2} & \frac{1}{2}  & 0 & 0 & 0
   \end{array} \right).
\end{equation}
Inserting Eq.~(\ref{PM1}) into Eq.~(\ref{Formular_PGF}), we obtain $\Phi(z)$ for this case. 
Then, the probability generating function of the FPT from one hub (i.e. hub $A$ ) to any of the other two hubs (i.e. hub $B$ or $C$) is
\begin{equation}\label{theta_t1}
  \theta(1,z)=\phi_{12}(z)+\phi_{13}(z)=\frac{-z}{z-2}.
\end{equation}

\paragraph*{Exact calculation of  $\phi_{C}(1,z)$ while hub $A$ is an absorbing node.}
Let two hub $A$  be the absorbing node. Then, all coordinates of the first row of the  transition probability matrix $\Pi$ are assigned $0$. Calculating $\Phi(z)$ from Eq.~(\ref{Formular_PGF}) 
we obtain
\begin{equation}\label{PHIC1}
  \phi_{C}(1,z)=\phi_{31}(z)=\frac{-z}{3z-4}.
\end{equation}

\paragraph*{Exact calculation of $\phi_{A}(1,z)$, $\phi_{B}(1,z)$ and $\phi_{C}(1,z)$ while nodes $D$, $E$ are  absorbing nodes.}
Let $D$ and $E$  be the absorbing nodes. Then all coordinates of the $4-th$ and $5-th$ row of the  transition probability matrix $\Pi$ are assigned $0$. Calculating $\Phi(z)$ from Eq.~(\ref{Formular_PGF}) 
we obtain
\begin{equation}\label{PHIA2}
  \phi_{A}(1,z)=\phi_{14}(z)+\phi_{15}(z)=\frac{z(z-2)}{3z-4},
\end{equation}
\begin{equation}\label{PHIB2}
  \phi_{B}(1,z)=\phi_{24}(z)+\phi_{25}(z)=\frac{-z}{3z-4},
\end{equation}
and
\begin{equation}\label{PHIC2}
  \phi_{C}(1,z)=\phi_{34}(z)+\phi_{35}(z)=\frac{-z}{3z-4}.
\end{equation}

\section{ Derivation of Eq.~(\ref{Rec_phi})}
 \label{App_phi}
Let  $\zeta_C(t)$ denote the FPT from $C$ to the absorbing domain.  Similar to Subsection~\ref{FPP0_2}, we can find independent random variables $N$, $\eta_1$, $\eta_2$, $\cdots$, such that
\begin{equation}
\label{pathzeta}
  \zeta_C(t)=\eta_1+\eta_2+\cdots+\eta_N,
\end{equation}
where $\eta_i$ ($i=1,2,\cdots$) are identically distributed random variables, each with probability generating function $\theta(t-1,z)$, and $N$ is the first-passage time from hub $C$ to  the absorbing domain on the simplified path including only nodes of level $1$. Therefore, the probability generating function of $N$ is $\phi_C(1,z)$.

Thus, we can obtain from Eqs.~(\ref{Sum_PGF}), (\ref{Sum_SN}) and (\ref{pathzeta}) that the probability generating function of $\zeta_C(t)$ satisfies
\begin{equation}
  \phi_C(t,z)=\phi_C(1,\theta(t-1,z)).
\end{equation}
\section{ Derivation of Eqs.~(\ref{g_Ck}), (\ref{T_Ck}) and (\ref{M2_Ck})}
 \label{App_g_C_k}
 Before proceeding, a remark is in order. For a random walker on $G(t)$, let $P_{x|y}(t,n)$ ($x,y=A$ or $B$) denote the splitting probability that the walker starting from the hub $C$ reaches the absorbing hub $x$  at time $n$ in the presence of another absorbing hub $y$. By symmetry, we have $P_{A|B}(t,n)=P_{B|A}(t,n)$. Therefore, we can  simply use  $Q(t,n)$ (defined in Sec.~IV A) to denote the sum of the two splitting probabilities, namely $Q(t,n) = P_{A|B}(t,n)+P_{B|A}(t,n) = 2P_{A|B}(t,n)$.

Now, we present the method to derive  Eqs.~(\ref{g_Ck}), (\ref{T_Ck}) and (\ref{M2_Ck}). Because any subunit $\Lambda_{k}$ ($k\geq0$) is a copy of the PSFW of generation $t-k$, the splitting probability that  a random walker starting from the node $C_k$ reaches the absorbing node $A_k$  at time $n$ in the presence of another absorbing node $B_k$ is equal to $\frac{1}{2}Q(t-k,n)$.  Noting that $\Lambda_{k}$ is connected to the two subunits of the PSFW by its two hubs ${A_k}$ and ${B_k}$, a random walker, starting from $C_k$,  must pass through either ${A_k}$ or ${B_k}$ before first reaching the absorbing domain which is not located in the starting subunit $\Lambda_{k}$.
 Thus,
     \begin{eqnarray}\label{P_Ck}
    P_{C_k}(t,n)&=&\frac{1}{2}\sum_{j=0}^{n}Q(t-k,j)P_{A_k}(t,n-j) \nonumber \\
    &&+\frac{1}{2}\sum_{j=0}^{n}Q(t-k,j)P_{B_k}(t,n-j).
  \end{eqnarray}

  Therefore, Eq.~(\ref{g_Ck}) is obtained according to the properties of generating function~\cite{Rudn04}. 

%
By taking the first order derivative on both sides of Eq.~(\ref{g_Ck})and posing $z=1$, we obtain  Eq.~(\ref{T_Ck}).

By taking the second order derivative on both sides of Eq.~(\ref{g_Ck}) and letting $z=1$, we obtain
 \begin{eqnarray}\label{T2_Ck}
      &&\left. \frac{\partial^2}{\partial z^2}\phi_{C_k}(t,z)\right|_{z=1}\nonumber\\
      &=&\frac{1}{2}\left[\left.\frac{\partial^2}{\partial z^2}\phi_{A_k}(t,z)\right|_{z=1}+\left.\frac{\partial^2}{\partial z^2}\phi_{B_k}(t,z)\right|_{z=1}\right] \nonumber\\
     &+&2^{t-k}[ \langle T_{A_k \rightarrow \mathfrak{D}} \rangle +  \langle T_{B_k \rightarrow \mathfrak{D}} \rangle]\nonumber\\
     &+&2^{2(t-k)+1}-2^{t-k+1}.
  \end{eqnarray}
  Therefore, Eq.~(\ref{M2_Ck}) is obtained from Eq.~(\ref{T2_Ck}) and the following equation.
  \begin{eqnarray}\label{aM2_Ck}
     \langle   T_{C_k \rightarrow \mathfrak{D} }^2 \rangle
      &=&\left. \frac{\partial^2}{\partial z^2}\phi_{C_k}(t,z)\right|_{z=1}+\langle   T_{C_k \rightarrow \mathfrak{D} } \rangle.
      \end{eqnarray}

\section{ Derivation of Eq.~(\ref{SSTC})}
 \label{sum_c1}

Note the following  mappings between hubs   of $\Lambda_{k-1}$ and hubs of $\Lambda_{k}$ (shown in FIG.~\ref{sub_k}). 
For any  $k>0$, Eq.~(\ref{T_Ck}) implies
\begin{equation}\label{RecT}
 \mathcal{T}^{(k)}=\mathcal{M}_{i_{k}}\mathcal{T}^{(k-1)}+\mathcal{V}^{(k)} \quad  i_{k}= 1, 2, 3,
\end{equation}
where
\begin{equation} \label{M12}
\mathcal{M}_{1}\!=\!
\left(                 
  \begin{array}{ccc}
   1 & 0 & 0 \\
   0 & 1 & 0 \\
   \frac{1}{2} & \frac{1}{2} & 0
  \end{array}
\right),
\,\,\,  \mathcal{M}_{2}\!=\!
\left(                 
  \begin{array}{ccc}
   1 & 0 & 0 \\
   0 & 0 & 1 \\
   \frac{1}{2} & 0 & \frac{1}{2}
  \end{array}
\right),
 \end{equation}
\begin{equation} \label{M3}
\mathcal{M}_{3}\!=\!
\left(                 
  \begin{array}{ccc}
   0 & 0 & 1 \\
   0 & 1 & 0 \\
   0 & \frac{1}{2} & \frac{1}{2}
  \end{array}
\right),
\,\,\,  \mathcal{V}^{(k)}=\left( \begin{array}{c} 0 \\ 0 \\ 2^{t-k}\end{array} \right).
\end{equation}
Using Eq.~(\ref{RecT}) recursively, for any $k>0$,
\begin{eqnarray}\label{TK1}
  \mathcal{T}^{(k)}
   &=&\mathcal{M}_{i_{k}}\mathcal{M}_{i_{k-1}}\cdots \mathcal{M}_{i_1}\mathcal{T}^{(0)}\nonumber \\
   & &+\sum_{l=1}^{k-1}\mathcal{M}_{i_k}\mathcal{M}_{i_{k-1}}\cdots \mathcal{M}_{i_{l+1}}\mathcal{V}^{(l)}+\mathcal{V}^{(k)}.
\end{eqnarray}
 As derived in Sec.~(\ref{FPP1_0}), $\langle T_{A \rightarrow \{A\}} \rangle=0$, $\langle T_{B \rightarrow \{A\}} \rangle=\langle T_{C \rightarrow \{A\}} \rangle=2^{t+1}$. Therefore  $T^{(0)}$ is known.
 Analogously to the method of Ref~\cite{Peng14b},  we obtain
\begin{eqnarray}\label{SST}
  \sum_{\{i_1,\cdots,i_k\}} \mathcal{T}^{(k)}
   &=& \mathcal{M}_{tot}^{k}\mathcal{T}^{(0)}+\sum_{l=1}^{k}3^{l}\mathcal{M}_{tot}^{k-l}\mathcal{V}^{(l)},
\end{eqnarray}
 where 
\begin{equation} \label{MTotal}
\mathcal{M}_{tot} = \mathcal{M}_{1}+\mathcal{M}_{1}+ \mathcal{M}_{3}.
\end{equation}
 Substituting  $\mathcal{M}_{i}$ from Eqs.~(\ref{M12}) and (\ref{M3}) in Eq.~(\ref{MTotal}), and orthogonal decomposing $\mathcal{M}_{tot}$, we obtain
 \begin{eqnarray}\label{MTS0}
 \mathcal{M}_{tot}^{k}\mathcal{T}^{(0)}
 &=&  \!\left( \!\begin{array}{c} 2^k+2\times 3^{k-1} \\ 2 \times 3^{k-1}-2^k  \\ 2 \times 3^{k-1} \end{array} \!\right)\! \times 2^{t+1},       \end{eqnarray}
and
\begin{eqnarray}\label{MTVT}
\sum_{l=1}^{k}3^{l}\mathcal{M}_{tot}^{k-l}\mathcal{V}^{(l)}
=3^{k-1}(2^t-2^{t-k}) \!\left( \!\begin{array}{c} 1 \\ 1 \\ 1\end{array} \!\right)\!     .
\end{eqnarray}
Inserting Eqs.~(\ref{MTS0}) and (\ref{MTVT})  into Eq.~(\ref{SST}) and calculating the sum of the three entries of $\sum_{\{i_1,\cdots,i_k\}}\mathcal{T}^{(k)}$, we obtain
 \begin{eqnarray}\label{PSSTC}
\Sigma_k=3^{k}(2^{t+2}+2^t-2^{t-k}).
  \end{eqnarray}

\section{ Derivation of Eq.~(\ref{SSTC2})}
 \label{sum_c2}

The method to derive Eq.~(\ref{SSTC2}) is similar to that of Eq.~(\ref{SSTC}).  The difference is that we must calculate $T^{(1)}$ directly in the case that nodes $D$, $E$ are absorbing nodes because absorbing nodes are located at nodes of level $1$. It is straightforward that  $\langle T_{D \rightarrow \{D,E\}} \rangle=\langle T_{E \rightarrow \{D,E\}} \rangle=0$. As derived in Sec.~(\ref{FPP2_0}), $\langle T_{A \rightarrow \{D,E\}} \rangle=3 \times 2^{t-1}$, $\langle T_{B \rightarrow \{D,E\}} \rangle=\langle T_{C \rightarrow \{D,E\}} \rangle=2^{t+1}$. Then,
 \begin{eqnarray}
 \langle T_{F \rightarrow \{D,E\}} \rangle
 &=&5\times2^{t-1}\nonumber.
   \end{eqnarray}
  Therefore  $T^{(1)}$ is known for all the cases of $i_1=1,2,3$. 
Similar to the derivation of Eq.~(\ref{TK1}), we have
\begin{eqnarray}\label{TK2}
  \mathcal{T}^{(k)}
   &=&\mathcal{M}_{i_{k}}\mathcal{M}_{i_{k-1}}\cdots \mathcal{M}_{i_2}\mathcal{T}^{(1)}\nonumber \\
   & &+\sum_{l=2}^{k-1}\mathcal{M}_{i_k}\mathcal{M}_{i_{k-1}}\cdots \mathcal{M}_{i_{l+1}}\mathcal{V}^{(l)}+\mathcal{V}^{(k)}.
\end{eqnarray}
Thus,
\begin{eqnarray}\label{SST2}
   \hspace{-5mm} \sum_{\{i_1,\cdots,i_k\}} \mathcal{T}^{(k)}
   = \mathcal{M}_{tot}^{k-1}\sum_{i_1=1}^3\mathcal{T}^{(1)}+\sum_{l=2}^{k}3^{l}\mathcal{M}_{tot}^{k-l}\mathcal{V}^{(l)}.
\end{eqnarray}

Similar to the derivation of Eq.~(\ref{PSSTC}), we have
 \begin{eqnarray}\label{PSSTC2}
\Sigma_k
      =3^{k}(2^{t+2}+2^t-2^{t-k}).
  \end{eqnarray}



\begin{thebibliography}{99}

\bibitem{LO93}
L. Lov\'{a}sz,  \emph{Combinatorics: Paul erd\"{o}s is eighty} (Keszthely, Hungary, 1993), vol. 2, issue 1, p. 1-46.

 \bibitem{Weiss94}
 G. Weiss, \emph{Aspects and Applications of the Random Walk} (Amsterdam, Netherlands: North-Holland, 1994).

 \bibitem{HaBe87}
 S. Havlin and D. ben-Avraham,  {Adv. Phys.} {\bf  36}, 695 (1987).


 \bibitem{Avraham_Havlin04}
D. ben-Avraham and S. Havlin,  \emph{ Diffusion and Reactions in
Fractals and Disordered Systems} (Cambridge University Press, Cambridge, UK, 2004).



\bibitem{MeKl00}
R. Metzler, J. Klafter, {Phys. Rep.}  {\bf 339}, 1 (2000).


 \bibitem{ChPe13}
 O. Chepizhko and F. Peruani,
{Phys. Rev. Lett.} {\bf111}, 160604 (2013).

 \bibitem{Redner07}
S. Redner, \emph{ A Guide to First-Passage Processes} (Cambridge University Press, Cambridge, UK, 2007).

\bibitem{MeyChe11}
B. Meyer, C. Chevalier, R. Voituriez,  and O.  O. B\'{e}nichou,
{Phys. Rev. E} {\bf83}, 051116 (2011).

 \bibitem{Condamin05}
S. Condamin, O. B¨¦nichou and M. Moreau, Phys. Rev. Lett. {\bf 95}, 260601 (2005).

\bibitem{HeMaKn04}
D. J. Heijs, V. A.  Malyshev, and J. Knoester,
J. Chem. Phys. {\bf 121}, 4884 (2004).


\bibitem{Ki58}
S. K. Kim,
J. Chem. Phys. {\bf 28}, 1057 (1958).

 \bibitem{Zhsh12}
Z. Z. Zhang, Y. B. Sheng, Z. Y. Hu, G. R. Chen,
Chaos {\bf 22}, 043129 (2012).

 \bibitem{CondaBe07}
 S. Condamin, O. B¨¦nichou, V. Tejedor, R. Voituriez, J. Klafter,
 Nature (London) {\bf 450}, 7166 (2007).




\bibitem{KoBa02}
J. J. Kozak and V. Balakrishnan  {  Phys. Rev. E }  {\bf65}, 021105 (2002).

\bibitem{BeTuKo10}
J. L. Bentz, J. W. Turner, J. J. Kozak  {Phys. Rev. E}  {\bf82}, 011137 (2010).

 \bibitem{ZhGuXi09}
 Z. Z. Zhang, J. H. Guan, W. L. Xie, Y. Qi and S. G. Zhou,
{Europhys. Lett.}  {\bf 86}, 10006 (2009).



\bibitem{ZhGaxie10}
 Z.Z. Zhang and S.Y.Gao, W. L. Xie,
 Chaos,  {\bf 20}, 043112 (2010).

\bibitem{CoMi10}
F. Comellas, A. Miralles,
{ Phys. Rev. E}  {\bf 81}, 061103 (2010).

  \bibitem{ZhZhGa10}
 Z. Z. Zhang, Y. Qi, S. G. Zhou, S. Y. Gao and J. H. Guan,  {Phys. Rev. E }  {\bf81}, 016114 (2010).

  \bibitem{LiWuZh11}
Y. Lin, B. Wu, and Z. Z. Zhang,
{Phys. Rev. E}  {\bf82}, 031140 (2010).

\bibitem{ZhLi11}
  Z. Z. Zhang , Y. Lin and Y. J. Ma,  { J. Phys. A: Math. Theor. }  {\bf44},  075102 (2011).

 \bibitem{ZhWu10}
  Z. Z. Zhang, B. Wu,  H. J. Zhang,  S. G. Zhou, J. H. Guan, and Z. G. Wang,
{Phys. Rev. E }  {\bf81},  031118 (2010).


\bibitem{ZhLiLin11}
 Z. Z. Zhang, X. T. Li, Y. Lin, G. R. Chen,
{J. Stat. Mech.: Theor. Exp.} P08013 (2011).

 \bibitem{Agl08}
E. Agliari,   {Phys. Rev. E}  {\bf77}, 011128 (2008).

 \bibitem{ZhYu09}
 Z. Z. Zhang, Y. Lin, S. G. Zhou, B. Wu, J. H. Guan ,
{ New J. Phys.}, {\bf11}, 103043 (2009).



\bibitem{WuLiZhCh12}
B. Wu, Y. Lin, Z. Z. Zhang, and G. R. Chen, {J. Chem. Phys.} {\bf 137}, 044903 (2012).

 \bibitem{AgCasCatSar15}
E. Agliari, F. Sartori, L. Cattivelli, and D. Cassi, Phys. Rev. E {\bf 91}, 052132 (2015).



\bibitem{AgBu09}
 E. Agliari and R. Burioni,
 {Phys. Rev. E}  {\bf 80}, 031125 (2009).

\bibitem{AgBuMa10}
 E. Agliari, R. Burioni  and A. Manzotti,
{Phys. Rev. E}  {\bf82}, 011118 (2010).


 \bibitem{Peng14b}
 J. H. Peng and  G. Xu,
{J. Chem. Phys.}  {\bf40},  134102 (2014).

 \bibitem{Peng14c}
 J. H. Peng, J. Xiong and G. Xu,
{Physica A} {\bf 407},  231 (2014).



\bibitem{BeChKl10}
O. B\'{e}nichou, C. Chevalier, J. Klafter, B. Meyer, and R. Voituriez,
{\it Nat. Chem.}  {\bf 2}, 472 (2010).

\bibitem{BenVo14}
O. B\'{e}nichou, R. Voituriez,
{Phys. Rep.} {\bf 539}, 4 (2014).



\bibitem{Dorogovtsev02}
S. N. Dorogovtsev, A. V. Goltsev, J. F. F. Mendes,
Phys. Rev. E
  {\bf
 65
 }, 066122 (2002).

\bibitem{zhang07b}
Z. Z. Zhang, L. L. Rong, S. G. Zhou,
 Physica A
{\bf
 377}%
, 329 (2007).

\bibitem{zhang07}
Z. Z. Zhang, S. G. Zhou, and L. C. Chen,
Eur. Phys. J. B
 {\bf
 58}
, 337 (2007).

\bibitem{zhang10}
Z. Z. Zhang, H. X. Liu, B. Wu, S. G. Zhou,
{Europhys. Lett.}  {\bf 90},
68002 (2010).

\bibitem{Bobe05}
E. M. Bollt, D. ben-Avraham,
New J. Phys.
{\bf
7}, 26 (2005).


\bibitem{ZhQiZh09}
 Z. Z. Zhang, Y. Qi, S. G. Zhou, W. L. Xie and J. H. Guan,
{Phys. Rev. E }  {\bf79}, 021127 (2009).

\bibitem{KahnRed89}
B. Kahng, S. Redner,
{J. Phys. A }  {\bf22}, 887 (1989).

\bibitem{Rudn04}
J. Rudnick£¬G. Gaspari,
\emph{ Elements of the Random Walk: An introduction for Advanced Students and Researchers} (Cambridge University Press, Cambridge, UK, 2004)

\bibitem{MeAgBeVo12}
B. Meyer, E. Agliari, O. B\'{e}nichou, and R. Voituriez,
{Phys. Rev. E }  {\bf85}, 026113 (2012).


\bibitem{RoHa07}
H. D. Rozenfeld, S. Havlin and D. ben-Avraham£¬
{New J. Phys.}, {\bf 9}  175 (2007).




\end{thebibliography}
\end{document}